\documentclass[12pt]{article}
\usepackage{graphicx}
\usepackage{epsfig}
\begin{document}
\baselineskip 18pt

\begin{center}
{\LARGE\bf  Polarization of $\Lambda^0$ hyperons in nucleus-nucleus collisions at high energies.}
\vspace{0.5cm}

{\Large\it  V.P.Ladygin{\footnote{E-mail: vladygin@jinr.ru}}, A.P.Jerusalimov, N.B.Ladygina }
\vspace{0.5cm}

{\Large Joint Institute for Nuclear Research, Dubna}
\vspace{0.5cm}

\end{center}

The measurement of $\Lambda^0$ hyperons polarization in nucleus-nucleus collisions
is considered as one of possible tools to study the phase transition.  
Fixed target and collider experiments are discussed for the case
of $\Lambda^0$'s production from $Au-Au$ central collisions at $\sqrt{s_{NN}}\sim$
of several GeV.
\vspace{0.5cm}


%\vspace{1cm}

\newpage

\section{Introduction}

The phase transition from ordinary nuclear matter to a quark gluon plasma (QGP)
should be observed when  conditions of sufficiently high baryonic densities and/or temperatures are achieved during the heavy-ion collisions.  
There is a number of experimental observables to identify 
this transition, such as strangeness enhancement, charmonium suppression,
event anisotropy, fluctuation in particles ratios, transverse momenta etc.
At the same time, it has also been recognized that no single signal
alone can provide clear evidence for the existence of the phase transition.
This stimulates the search for new probes to investigate
the properties of hot and dense matter produced in heavy-ion collisions at
high energies.

One of the new signature can be the change of in the polarization properties
of the secondary particles in the nucleus-nucleus collisions compared to the
nucleon-nucleon collisions.  A number of polarization observables have been proposed as 
a possible signature of phase transition, namely, 
decreasing of the ${\Lambda^0}$ transverse polarization in central collisions 
\cite{lamb0,ayala},  non-zero $\bar{\Lambda}^0$ longitudinal polarization \cite{a_lam1,a_lam2}, non-zero $J/{\Psi}$ polarization at low $p_T$ \cite{ioffe},
anisotropy in di-electron production from vector mesons decay \cite{brat} (and references therein),  global hyperon polarization \cite{global_th} and
spin-alignment of vector mesons \cite{global_th1} in non-central events  etc.

The strangeness enhancement is considered as one of the
signal of the phase transition from ordinary matter to quark-gluon plasma (QGP).
The strong strangeness enhancement
at $\sim$30 GeV$\cdot$A as well as the change of the slope of the strange particles
spectra have been firstly observed by the NA49 collaboration both in kaon \cite{NA49_K}
and hyperon \cite{NA49_lambda} data. In this respect the study of the polarization
properties of the hyperons at the energies $\sqrt{s_{NN}}$ of several GeV is of greate 
interest.

However, in spite of the large amount of the data on the $\Lambda^0$ production
obtained at GSI \cite{gsi}, AGS \cite{eos,ags,ags1},  
SPS \cite{NA49_lambda} and RHIC \cite{ph1}-\cite{star_lambda4}, 
there are only few results on 
the $\Lambda^0$ polarization in nucleus-nucleus interaction 
\cite{har,anik,bnl,global_STAR}.

The goal of present article is to summarize the existing experimental data and
to discuss the possibility of the $\Lambda^0$ polarization measurements in
nucleus-nucleus collisions at
$\sqrt{s_{NN}}$ of several GeV at future setups.

\section{$\Lambda^0$ polarization in proton-induced reactions}

Since the first observation of large transverse polarization of $\Lambda^0$ hyperon 
\cite{lesnik,fermilab}
in inclusive nucleon-nucleon and nucleon-nuclei interactions, there has been permanent 
interest to the origin of this effect.
A significant amount of the data on $\Lambda^0$ polarization 
has been  accumulated at different nucleon initial energies 
between 6 and $\sim$2000 GeV \cite{Lambda_pA}.
All these measurements indicated that the high energy hyperon polarization is a function  
only of the hyperon transverse momentum $p_T$ and the fraction of the beam energy carried 
by the final state hyperon $x_F$. The magnitude of the polarization at fixed $x_F$ rises
with $p_T$ to a  plateau at about 1 GeV/$c$, and the size of the plateau increases monotonically
with $x_F$. A minor dependence of the $\Lambda$ polarization  on the A-value of the target has been observed.

Most of the models explain the $\Lambda^0$
polarization by the recombination of a polarized $s$- quark with an unpolarized 
($ud$) spectator diquark. 
In this case the spin of $\Lambda^0$ is carried by the produced
$s$- quark and that the $u$ and $d$ quarks can be thought 
of as being coupled into a diquark with zero total angular momentum
and isospin.

In the framework of the semiclassical recombination model 
of DeGrand-Miettinen \cite{DMM} the $\Lambda^0$ polarization is defined by the slow
$s$- quark spin precession. According to this model $\Lambda^0$ polarization is
expressed as following 
\begin{eqnarray}
{\cal P}_{\Lambda^0} = 
-\frac{12}{\Delta x_0 M^2}\cdot \frac{1-3\xi(x)}{[1+3\xi(x)]^2}\cdot p_T,
\end{eqnarray}
where
\begin{eqnarray}
M^2=  \frac{m_D^2+p_{TD}^2}{1-\xi(x)}+ \frac{m_s^2+p_{Ts}^2}{\xi(x)}-m^2_{\Lambda^0}-p_T^2
\end{eqnarray}
with $m_D,p_{TD}$ ($m_s,p_{Ts}$) the mass and transverse momentum of the 
$ud$ diquark ($s$ quark), $m_{\Lambda^0}$ and $p_T$ 
the mass and transverse momentum of the $\Lambda^0$, $\Delta x_0$ a distance scale and
$\xi(x)=x_s/x$ the ratio of the longitudinal momentum of the $s$ quark to the longitudinal
momentum fraction of the $\Lambda^0$ with respect to the beam proton.

The parametrization taken in the form
\begin{eqnarray}
\xi(x) = \frac{1}{3}(1-x)+ 0.1 x
\end{eqnarray}
gives a quantitative description for the $\Lambda^0$ polarization in nucleon-nucleus  interaction a wide 
energy range \cite{lesnik,fermilab,Lambda_pA}.

\section{$\Lambda^0$ polarization in heavy-ion collisions}

In the case of relativistic nucleus-nucleus collisions, the expectation
is that, $\Lambda^0$'s coming from the region where the critical
density for QGP formation has been achieved, are produced through the
coalescence of independent slow sea $u$, $d$ and $s$ quarks.
Therefore, the plasma creates $\Lambda^0$ with zero polarization
\cite{ayala}.  
Finally, in the case of QGP formation 
the depolarization effect compare to proton-induced reaction
should be observed. 
The total cross section of $\Lambda^0$ production versus impact parameter $b$ 
is the sum of the cross sections for the peripherical collisions and QGP regions \cite{ayala}
\begin{eqnarray}
\frac{d^2\sigma_{\Lambda^0}}{d^2b} =
\frac{d^2\sigma^{PER}_{\Lambda^0}}{d^2b}+\frac{d^2\sigma^{QGP}_{\Lambda^0}}{d^2b}.
\end{eqnarray}
Since $\Lambda^0$'s from QGP zone have a zero polarization one can write:
\begin{eqnarray}
{\cal{P}}_{\Lambda^0} &=& \frac{{\cal{P}}^{PER}_{\Lambda^0}}{1+f(b)},\\
f(b) &= & \frac{d^2\sigma^{QGP}_{\Lambda^0}}{d^2b}/
\frac{d^2\sigma^{PER}_{\Lambda^0}}{d^2b},
\end{eqnarray}
where $f(b)$ is defined as the ratio of the differential cross sections
for  $\Lambda^0$ production  from the QGP and recombination processes.

The behavior of $f(b)$ as a function of impact parameter $b$ for RHIC energies 
is shown in Fig.1 \cite{ayala}. One  can 
expect up to 30--40\% of the depolarization effect for the central events.
It is shown that the measurable depolarization effect 
can be observed at the $\Lambda^0$ transverse momenta $p_T\ge$0.6--0.8~GeV/$c$. 

The measurements of $\Lambda^0$ transverse polarization in nucleus-nucleus collisions
\cite{har,anik,bnl} have shown that 
the produced $\Lambda^0$'s are still polarized at freeze-out that means 
the spin direction is only little affected by the rescattering phase after hadronization.
However, there was no scan versus the centrality in this experiment.

Another novel phenomena is so called global polarization \cite{global_th,global_th1}.
System in the noncentral collisions have large 
orbital angular momentum, which manifests in the
polarization of secondary particles along the direction of
the  system angular  momentum.

Parton interaction in noncentral relativistic
nucleus-nucleus collisions leads  to the global polarization
of the produced quarks. This value for RHIC energies can be about 30\% and 
would lead to the global polarization of hyperons \cite{global_th}.
More realistic calculations  \cite{global_th2} within a model based
on the hard termal loop gluon propagator predict the value of hyperon polarization
to be in the range from -0.03 to 0.15 depending on the temperature of
the QGP formed. Recent calculations \cite{global_th3,global_th4} also explain
the small value of the global $\Lambda^0$ polarization observed at RHIC \cite{global_STAR}.

\section{Method of the $\Lambda^0$ polarization measurement}

The $\Lambda^0$ hyperon has spin $1/2$. It decays to $p+\pi^-$ with
64\% branching ratio. The analysis of the angular distribution
can be used to obtain the polarization value.
The main method to measure $\Lambda^0$ polarization in nucleus-nucleus collisions is the
measurement of the emission angle distribution of the decay proton with respect to
the system orbital momentum $\bf L$.

The definition of the different angles for  $\Lambda^0$- hyperon decay is given
in Fig.\ref{fig:lambda_fig2}. Here
$\bf x$, $\bf y$, $\bf z$ are the laboratory frame axes.  The reaction plane is defined
by the vectors of the initial beam direction $\bf z$ and impact parameter $\bf b$.
$\bf L$ is the system orbital momentum normal to the reaction plane.
$\bf p^*$ and $\theta^*$ are the proton three-momentum 
and the angle between the system orbital momentum $\bf L$ and the
three-momentum of the proton
in the $\Lambda^0$ rest frame, respectively. 
$\Psi_{RP}$ is the reaction plane angle.

The decay proton distribution as a function of emission angle $\theta^*$ 
is expressed as
\begin{eqnarray}
\frac{dN}{dcos\theta^*} = N_0(cos\theta^*)\large ( 1 + \alpha {\cal P}_{\Lambda} cos\theta^* \large ), 
\end{eqnarray}
where ${\cal P}_\Lambda$ is the value of $\Lambda^0$ polarization, $\theta^*$ is the angle
in the $\Lambda^0$ rest frame between the system orbital momentum $\bf L$ which is normal
to the reaction plane and 
the decay proton,  $\alpha$ is an $s-p$-wave 
interference term factor which is measured to be 0.642,
$N_0(cos\theta^*)$ is the constant corrected for the geometry, efficiency etc.
The polarization is defined as the slope of the ${dN}/{dcos\theta^*}$ vs 
$cos\theta^*$. 

The global polarization of the hyperons  measured at RHIC 
\cite{global_STAR} is defined as the polarization ${\cal P}_\Lambda$ averaged
over the relative azimuthal angle
\begin{eqnarray}
{\cal P}_{\Lambda}^G = \frac{3}{\alpha}<cos\theta^*>.
\end{eqnarray}
One can write global polarization in terms of the reaction plane angle $\Psi_{RP}$ and
the azimuthal angle $\phi^*_p$ of the proton three-momentum in the 
$\Lambda^0$ rest frame  using the relation  among the angles shown in 
Fig.\ref{fig:lambda_fig2} which is $cos\theta^*=sin\theta^*_psin(\phi^*_p-\Psi_{RP})$ 
and integrating over the angle $\theta^*_p$
\begin{eqnarray}
\label{Pglobal}
{\cal P}_{\Lambda}^G = \frac{8}{\pi\alpha}<sin(\phi^*_p-\Psi_{RP})>.
\end{eqnarray}
The expression (\ref{Pglobal}) is similar to that used for
the directed flow measurements.

\section{Feasibility of $\Lambda^0$ polarization measurement at high energies}

From UrQMD transport model
one can expect production of approximately 20 $\Lambda^0$'s decaying into 
$p\pi^-$ pair in the central $Au-Au$ collisions at 25~GeV$\cdot$A.

The important feature of CBM setup \cite{CBM_experiment}
is quite large acceptance for the detection of the both
protons and pions.
Therefore, it is possible to detect the $\Lambda^0$- hyperons
with large transverse momenta $p_T$, where
maximal effect in polarization properties is expected. 

Silicon Tracking System (STS)  consisted of 7 or 8 stations can 
provide the good tracking with the expected good enough efficiency.
Since $\Lambda^0$ is weakly decaying particle ($c\tau$=7.89 cm),
its decay vertex is well separated 
from the primary vertex. However, it is necessary
to provide high efficiency of the secondary tracks finding. 
The selection of $\Lambda^0$- hyperons at CBM can be done without 
the secondary particles identification to increase the acceptance.
In this case the 
main source of the background is $K_S^0$ meson decaying into $\pi^+\pi^-$ pair.

The detailed feasibility study of the $\Lambda^0$- hyperon production at CBM \cite{CBM_kryshen} has been 
performed using track reconstruction based on the cellular automata method 
\cite{CBM_experiment}. 
The mass and longitudinal secondary vertex resolutions achieved were $M_{\Lambda}\sim 0.85$~MeV/$c^2$ and 0.8~mm, respectively. 

An alternative approach has been developed  in ref.\cite{JAP}. 
Special $V^0$- particles ($\Lambda^0$ and $K^0_S$) finder has been developed 
using track reconstruction based on the conformal mapping and approximate solution of motion equation \cite{CM}. The precision of secondary vertex reconstruction was $\sim 150~\mu$m
and $\sim 100~\mu$m for $\Lambda^0$ and $K^0_S$, respectively. The simple $V^0$- fit used
the energy and momentum conservation laws only
\begin{eqnarray}
E_{V^0} &=& E^++E^-\nonumber\\
\vec{p}_{V^0} &=& \vec{p}^{~+}~+~\vec{p}^{~-}
\end{eqnarray}
with the fixed values of the $V^0$- particles masses \cite{LVP}.
The purity of $\Lambda^0$- hyperons selection with simple $V^0$- fit  
is shown  in Fig.3 as a function of the longitudinal vertex position. 
One can see that the purity of $\Lambda^0$'s selection  is about 0.97 for
the decay longitudinal vertex position larger than 5~mm.

The $y-p_T$ acceptance for reconstructed $\Lambda^0$'s from $Au-Au$ central collisions
at 25~GeV$\cdot$A for the conditions of CBM experiment is shown in Fig.4.

However, the measurement of the $\Lambda^0$ polarization requires good knowledge of 
the kinematical parameters of the  $\Lambda^0$, as well as the reaction plane and 
centrality value. The latest two parameters at CBM can be measured by zero degree calorimeter \cite{CBM_PSD} with good precision. 
The $V^0$- fitter \cite{JAP} improves significantly the parameters of the $\Lambda^0$ hyperons.
The momentum and angular resolutions achieved are $\delta~p/p\sim$0.37\%, 
$\sigma_\beta\sim$0.13~mrad and $\sigma_{\tan{\alpha}}\sim$0.13, respectively.
The precision of secondary vertex reconstruction was better than $\sim 70~\mu$m.
The distributions on the reconstructed momentum, angles and vertex position for 
$\Lambda^0$'s are given in Fig.5.

The collider mode provides more symmetrical acceptance for $\Lambda^0$ hyperons,
that is significant for the proper reconstruction of the polarization.
The  $y-p_T$ ideal acceptance for $\Lambda^0$'s 
from $Au-Au$ central collisions at $\sqrt{s_{NN}}$= 7.1~GeV for the collider mode
(assuming the axial symmetry of MPD  \cite{MPD_experiment}) 
is show in Fig.6. Since no particle identification is required
the transverse momenta $p_T$ achieved can be large enough to measure sizable polarization
effects for $\Lambda^0$'s  production. The reaction plane and 
impact parameter at MPD \cite{MPD_experiment} can be measured with zero degree
calorimeter assumed to be the same as at CBM setup or via charged particles
energy flow reconstructed in the time-projection chamber.

\section{Conclusions}

The measurement of the $\Lambda^0$ polarization in nucleus-nucleus collisions
at the energies of FAIR and NICA can be a new important tool in addition 
the traditional ones to study the phase transition.

It is shown the feasibility of the $\Lambda^0$ selection in $Au-Au$ 
collisions at 25~GeV$\cdot$A for the fixed target experiment \cite{CBM_experiment}.
The study of the polarization effects in hyperon production
for the collider mode \cite{MPD_experiment} at the same energies has certain
advantages and could be promising.  

Authors thank Prof.~A.I.~Malakhov for the stimulating support of this work 
and O.V.~Rogachevsky for the help in the simulation.
This work has been supported in part by the Russian Foundation for Basic Researches
under grant No.07-02-00102a.

\newpage

\begin{figure}[htbp]
 \centering
  \resizebox{10cm}{!}{\includegraphics{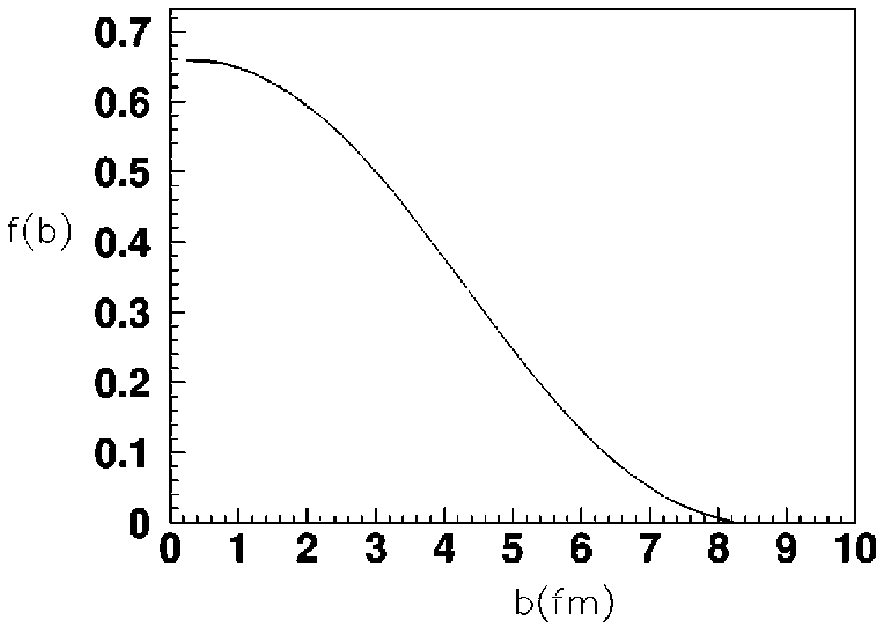}}
\caption{The ratio of the differential cross sections
for  $\Lambda^0$ production  from the QGP and recombination processes \cite{ayala}.}
\label{fig:lambda_fig1}
\end{figure}

\newpage

\begin{figure}[htbp]
 \centering
  \resizebox{10cm}{!}{\includegraphics{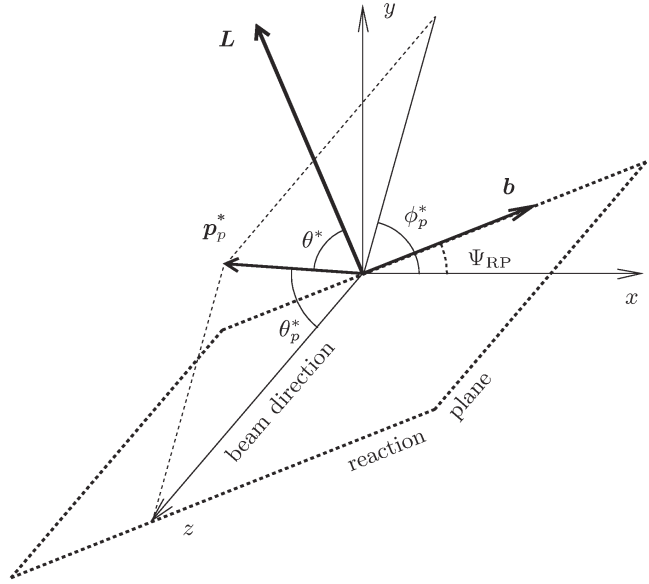}}
\caption{The definition of the different angles for  $\Lambda^0$- hyperon decay.
$\bf x$, $\bf y$, $\bf z$ are the laboratory frame axes.  The reaction plane is defined
by the vectors of the initial beam $\bf z$ and impact parameter $\bf b$.
$\bf L$ is the system orbital momentum normal to the reaction plane.
$\bf p^*$ and $\theta^*$ are the proton three-momentum 
and the angle between the system orbital momentum $\bf L$ and the
three-momentum of the proton
in the $\Lambda^0$ rest frame, respectively. 
$\Psi_{RP}$ is the reaction plane angle.
The picture is taken from ref.\cite{global_STAR}.}
\label{fig:lambda_fig2}
\end{figure}

\newpage

\begin{figure}[htbp]
 \centering
  \resizebox{10cm}{!}{\includegraphics{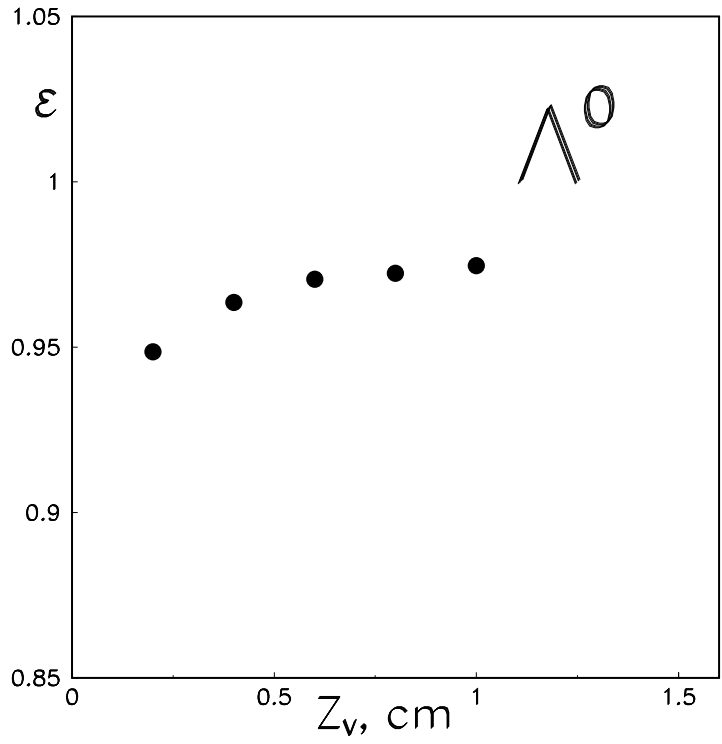}}
\caption{The purity of $\Lambda^0$- hyperons selection with simple kinematical $V^0$-
fit as a function of the 
longitudinal secondary vertex position.
}
\label{fig:lambda_fig3}
\end{figure}

\newpage

\begin{figure}[htbp]
 \centering
  \resizebox{10cm}{!}{\includegraphics{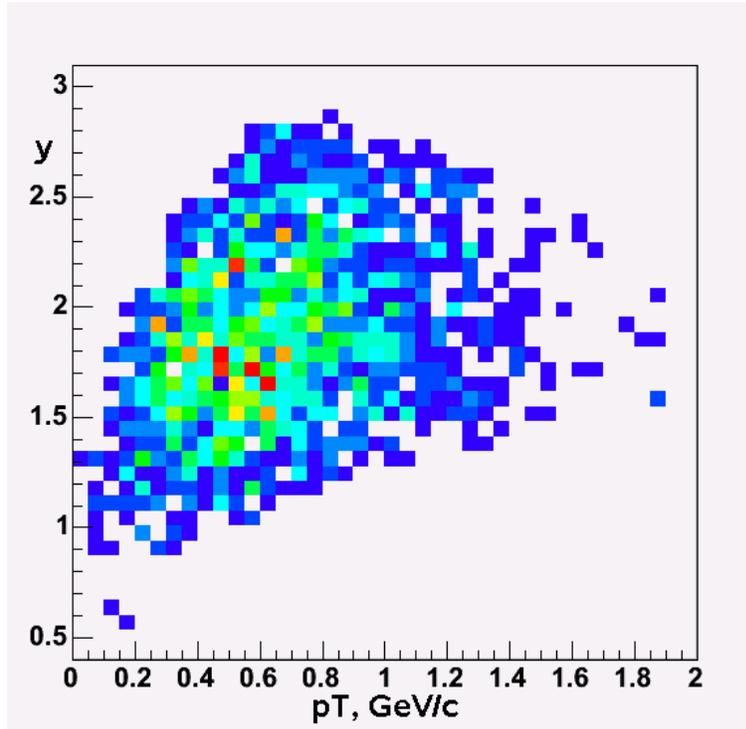}}
\caption{The $y-p_T$ acceptance for reconstructed $\Lambda^0$'s from $Au-Au$ central collisions
at 25~GeV$\cdot$A for the conditions of CBM experiment. 
}
\label{fig:lambda_fig4}
\end{figure}

\newpage

\begin{figure}[htbp]
\begin{minipage}[t]{0.47\textwidth}
 \centering
  \resizebox{8cm}{!}{\includegraphics{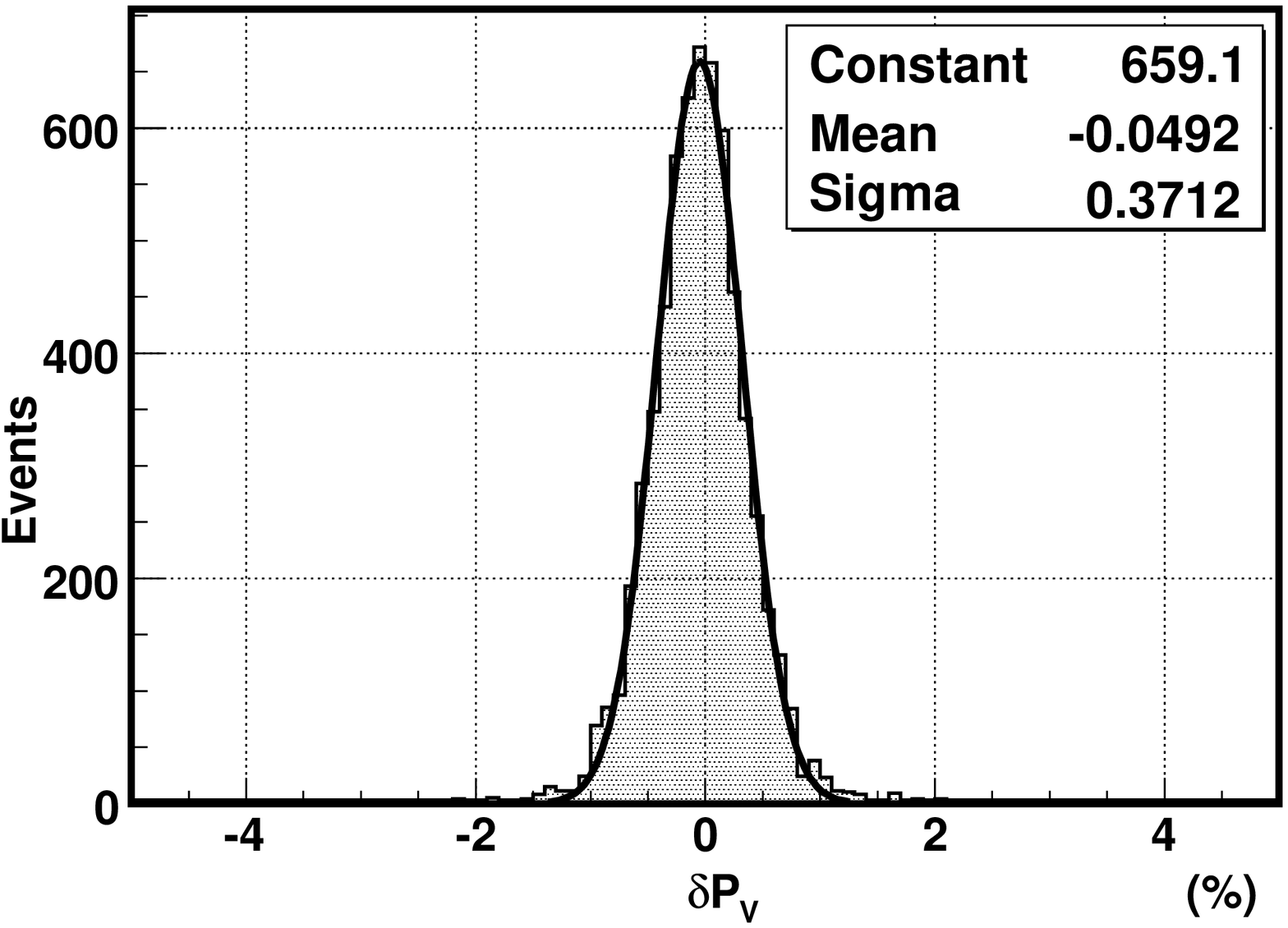}}
\label{fig:lambda_fig5a}

\end{minipage}\hfill
\begin{minipage}[t]{0.47\textwidth}
 \centering
  \resizebox{8cm}{!}{\includegraphics{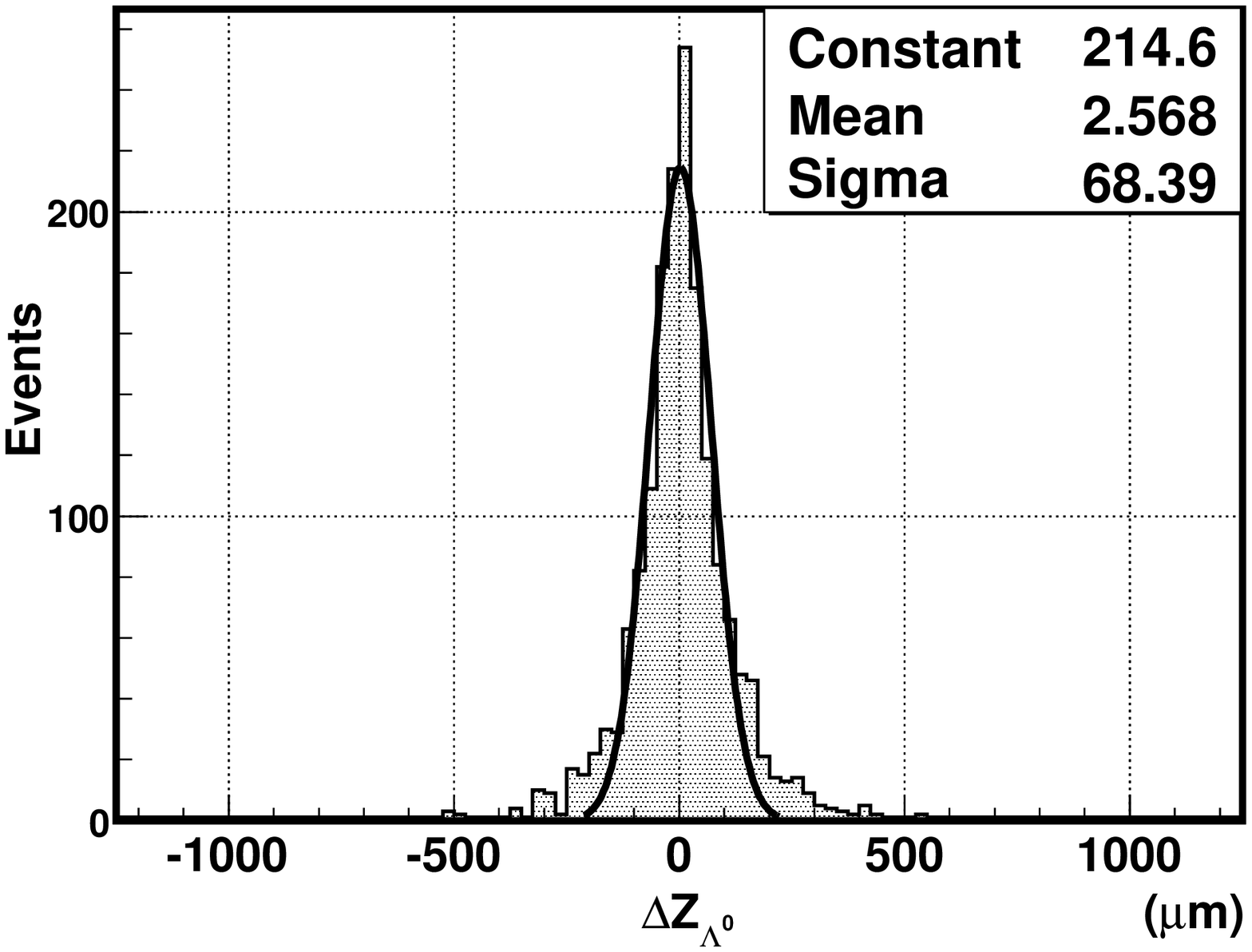}}
\label{fig:lambda_fig5b}
\end{minipage}

\begin{minipage}[t]{0.47\textwidth}
 \centering
  \resizebox{8cm}{!}{\includegraphics{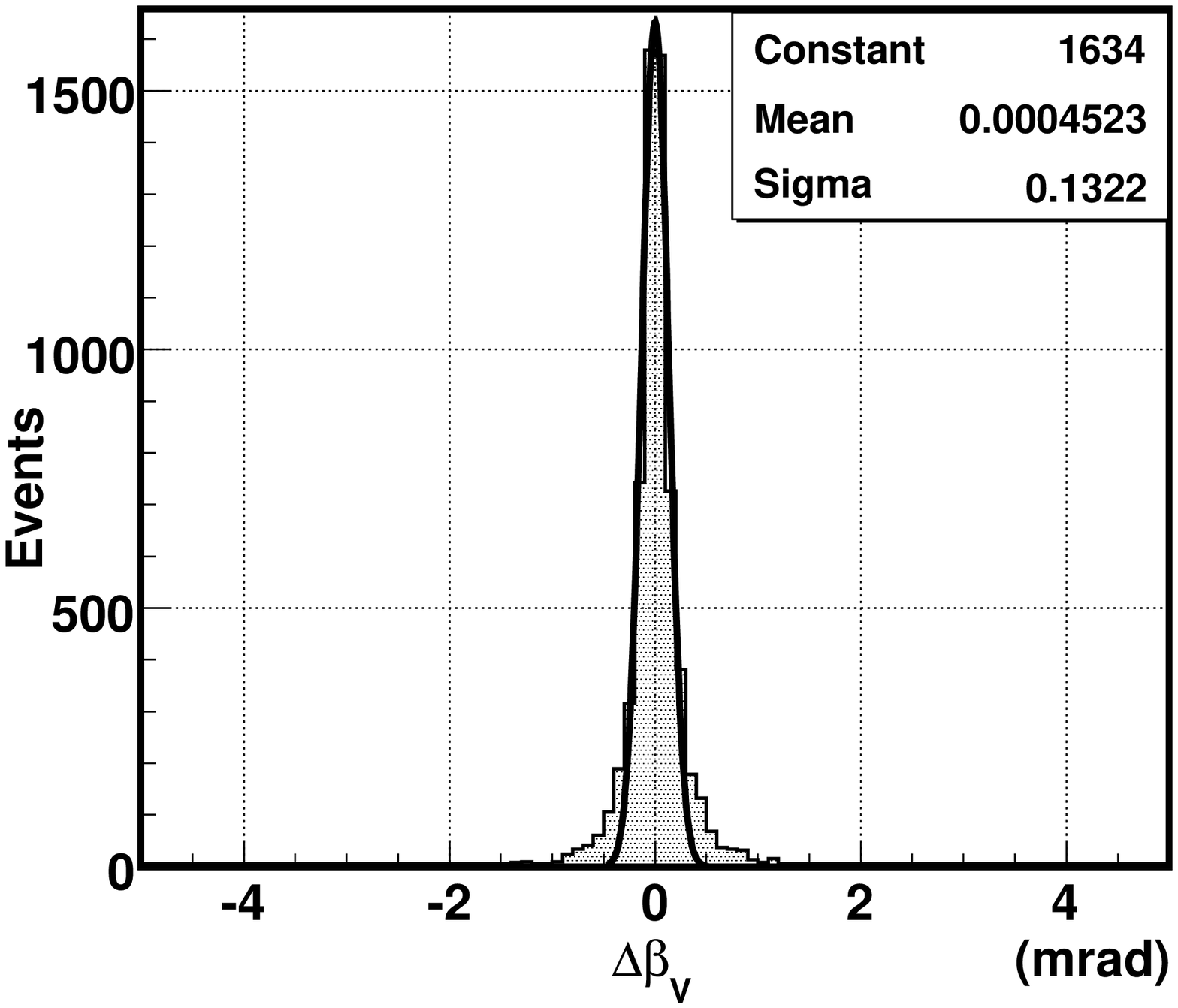}}
\label{fig:lambda_fig5c}

\end{minipage}\hfill
\begin{minipage}[t]{0.47\textwidth}
 \centering
  \resizebox{8cm}{!}{\includegraphics{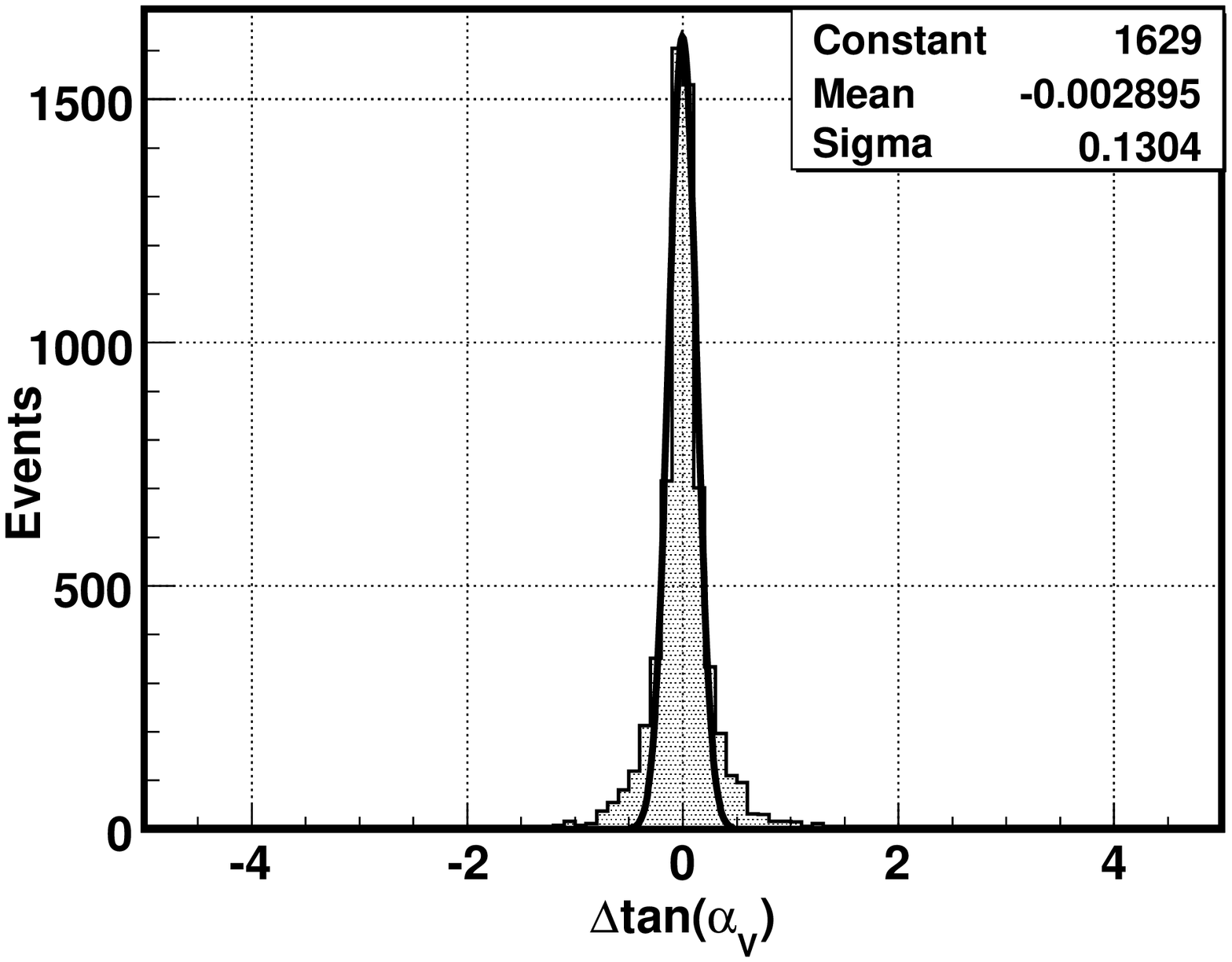}}
\label{fig:lambda_fig5d}
\end{minipage}
\caption{The momentum, vertex position  and angles reconstruction precisions for 
$\Lambda^0$'s from $Au-Au$ central collisions at 25~GeV$\cdot$A \cite{JAP}. 
} 
\end{figure}

\newpage

\begin{figure}[htbp]
 \centering
  \resizebox{10cm}{!}{\includegraphics{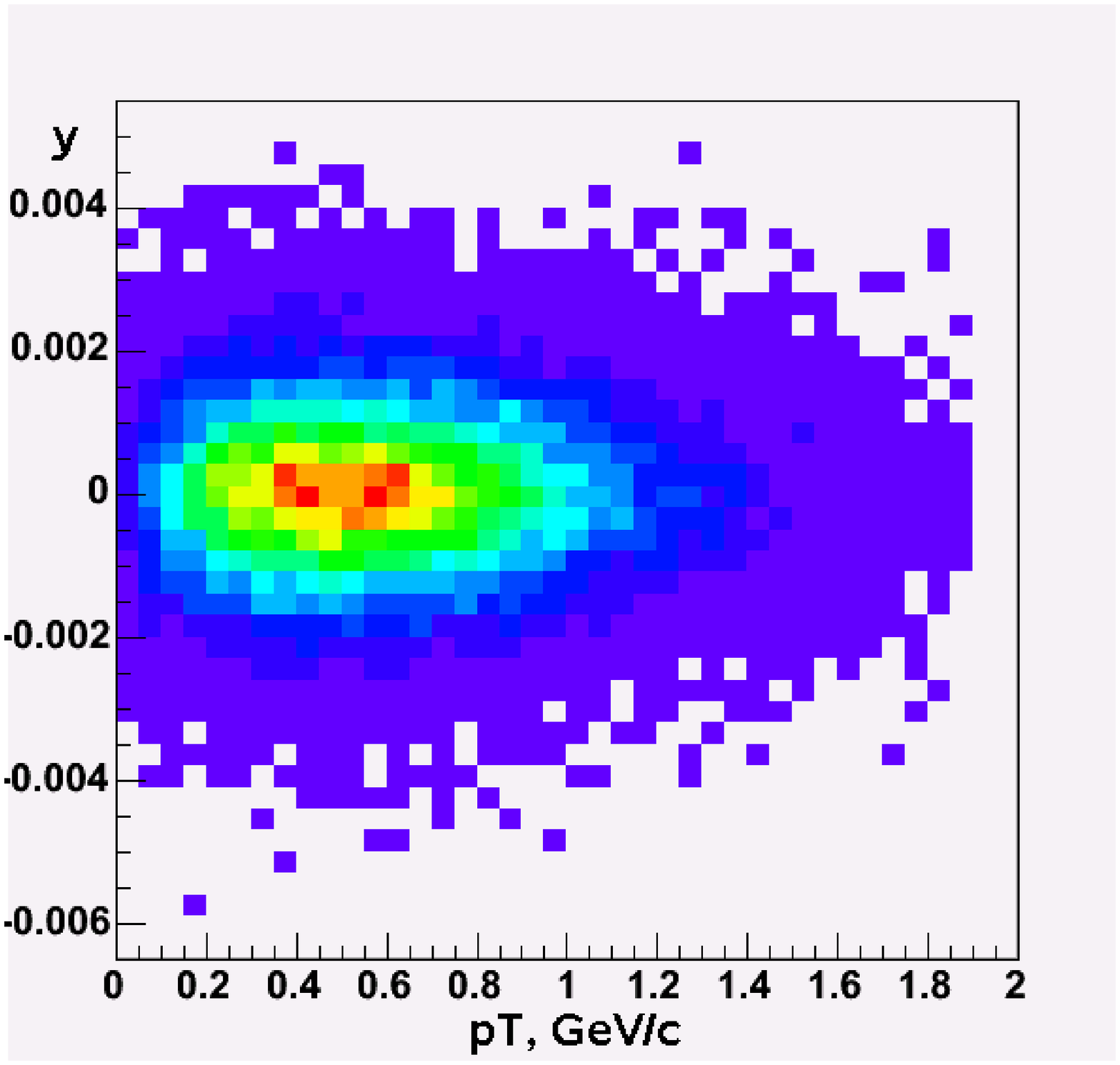}}
\caption{The $y-p_T$ acceptance for  $\Lambda^0$'s from $Au-Au$ central collisions at $\sqrt{s_{NN}}$= 7.1~GeV for the collider mode. 
}
\label{fig:lambda_fig6}
\end{figure}


\begin{thebibliography}{99}
\bibitem{lamb0} {\it Panagiotou~A.D.}// 
Phys.Rev. 1986. {V.C33.} P.1999.
\bibitem{ayala} {\it Ayala~A., Cuautle~E., Herrera~G. and 
Montano~L.M.}// Phys.Rev. 2002. {V.C65.} P.024902.
\bibitem{a_lam1} {\it Jacob~M.}// Z.Phys. 1988. {V.C38.} P.273.
\bibitem{a_lam2} {\it Herrera~G., Magnin~G. and Montano~L.M.}// 
Eur.Phys.J. 2005. {V.C39.} P.95. 
\bibitem{ioffe} {\it Ioffe~B.L. and Kharzeev~D.E.}// Phys.Rev. 2003. {V.C68.} P.061902(R).
\bibitem{brat} {\it Bratkovskaya~E.L., Cassing~W. and Mosel~U.}// 
Z.Phys. 1997. {V.C75.} P.119.
\bibitem{global_th} {\it Liang~Z.-T. and Wang~X.-N.}//  
Phys.Rev.Lett. 2005. {V.94}. P.102301.
\bibitem{global_th1} {\it Liang~Z.-T. and Wang~X.-N.}//  
Phys.Lett. 2005. {V.B629}. P.20.
\bibitem{NA49_K} {\it Afanasiev~S.V. et al.} Phys.Rev. 2002. {V.C66.} P.054902.
\bibitem{NA49_lambda} {\it Anticic~T. et al.} Phys.Rev.Lett. 2004. {V.93.} P.022302.
\bibitem{gsi} {\it Merschmeyer~M. et al.}// Phys.Rev. 2007. {V.C76.} P.024906.
\bibitem{eos} {\it Justice~M. et al.}// Phys.Lett. 1998.{V.B440.} P.12.
\bibitem{ags} {\it Albergo~S. et al.}// Phys.Rev.Lett. 2002. {V.88}. P.062301.
\bibitem{ags1} {\it Barrette~J. et al.}// Phys.Rev. 2000. {V.C63.} P.014902.
\bibitem{ph1} {\it Adcox~K. et al.}// Phys.Rev.Lett. 2002. {V.89.} P.092302.
\bibitem{star_lambda1} {\it Adler~C. et al.}// Phys.Rev.Lett. 2002. {V.89.} P.092301.
\bibitem{star_lambda2} {\it Abelev~B.I. et al.}// Phys.Rev.Lett. 2006. {V.97.} P.132303.
\bibitem{star_lambda3}{\it Adams~J. et al.}// Phys.Rev.Lett. 2007. {V.98.} P.062301.
\bibitem{star_lambda4} {\it Abelev~B.I. et al.}// arXiv:0705.2511v1 [nucl-ex].
\bibitem{har} {\it Harris~J. et al.}// Phys.Rev.Lett. 1981. {V.47.} P.229.
\bibitem{anik} {\it Anikina~M. et al.}// Z.Phys. 1984. {V.C25.} P.1.
\bibitem{bnl} {\it Bellwied~R. et al.}// Nucl.Phys. 2002. {V.A698.} P.499c. 
\bibitem{global_STAR} {\it Abelev~B.I. et al.}// Phys.Rev. 2007. {V.C76.} P.024915.
\bibitem{lesnik} {\it Lesnik~A. et al.}// Phys.Rev.Lett. 1975. {V.35.} P.770;
\bibitem{fermilab} {\it Bunce~G. et al.}// Phys.Rev.Lett. 1976. {V.36.} P.1113;
\bibitem{Lambda_pA} {\it Heller~K. et al.}// Phys.Lett. 1977. {V.68B.} P.480; 
Phys.Rev.Lett. 1978. {V.41.} P.607; ibid. 1983. {V.51.} 2025;

{\it Abe~F. et al.}// Phys.Rev.Lett. 1983. V.50. P.1102;
Phys.Rev. 1986. V.D34. P.1950;

{\it Bonner~B.E. et al.}// Phys.Rev. 1988. V.D38. P.729;

{\it A.M.~Smith et al.}// Phys.Lett. 1987. V.B185. P.209;

{\it B.~Lundberg et al.}// Phys.Rev. 1989. V.D40. P.3557;

{\it Henkes~T. et al.}// Phys.Lett. 1992. V.B283.  P.152;

{\it Ramberg~E.J. et al.}// Phys.Lett. 1994. V.B338. P.403;

{\it Fanti~V. et al.}// Eur.Phys.J. 1999. V.C6. P.265;

{\it Aleev~A.N. et al.}// Eur.Phys.J. 2000. V.C13. P.427.
%\bibitem{heller} {\it Heller~K. et al.}// Phys.Lett. 1977. {V.68B.} P.480; 
%Phys.Rev.Lett. 1978. {V.41.} P.607; ibid. 1983. {V.51.} 2025.  
%\bibitem{abe} {\it Abe~F. et al.}// Phys.Rev.Lett. 1983. V.50. P.1102;
%Phys.Rev. 1986. V.D34. P.1950.
%\bibitem{bonner} {\it Bonner~B.E. et al.}// Phys.Rev. 1988. V.D38. P.729.
%\bibitem{smith} {\it A.M.~Smith et al.}// Phys.Lett. 1987. V.B185. P.209.
%\bibitem{lundberg} {\it B.~Lundberg et al.}// Phys.Rev. 1989. V.D40. P.3557.
%\bibitem{r608} {\it Henkes~T. et al.}// Phys.Lett. 1992. V.B283.  P.152.
%\bibitem{ramberg} {\it Ramberg~E.J. et al.}// Phys.Lett. 1994. V.B338. P.403.
%\bibitem{na48} {\it Fanti~V. et al.}// Eur.Phys.J. 1999. V.C6. P.265.
%\bibitem{excharm} {\it Aleev~A.N. et al.}// Eur.Phys.J. 2000. V.C13. P.427.
%\bibitem{felix} J.~F\'elix et al., Phys.Rev.Lett.88 (2002) 061801.
\bibitem{DMM} {\it DeGrand~T.A. and Miettinen~H.I.}// Phys.Rev. 1981. V.D23. P.1227;
Phys.Rev. 1981. V.D24. P.2419; Phys.Rev. 1985. V.D31. P.661(E);

{\it DeGrand~T.A., Markkanen~J. and Miettinen~H.I.}// Phys.Rev. 1985. V.D32. P.2445.
\bibitem{global_th2} {\it Liang~Z.-T.}// J.Phys.G: Nucl.Part.Phys. 2007. {V.34.} P.S323.
\bibitem{global_th3} {\it  Gao~J.-H. et al.}// arXiv:0710.2943 [nucl-th].
\bibitem{global_th4} {\it Barros~Jr.~C.C. and Hama~Y.}// arXiv:0712.3447 [nucl-th].
\bibitem{CBM_experiment} {\it CBM Experiment}// Technical Status Report. 2005.
(http://www.gsi.de/onTEAM/dokumente/public/DOC-2005-Feb-447e.html).
\bibitem{CBM_kryshen} {\it Kryshen~E. and Berdnikov~Y.}// CBM-PHYS-note-2005-002. 2005.
\bibitem{JAP} {\it Jerusalimov~A.P. et al.}// Talk at 9-th CBM-Collaboration meeting,
28 February -2 March. 2007; http://www.gsi.de/documents/DOC-2007-Mar-46-1.pdf. 
\bibitem{CM} {\it Rogachevsky~O.V. and Jerusalimov~A.P.}//
Talk at the 9-th CBM-Collaboration meeting,
28 February -2 March. 2007; http://www.gsi.de/documents/DOC-2007-Mar-45-1.pdf.
\bibitem{LVP} {\it Ladygin~V.P. et al.}//Talk at 9-th CBM-Collaboration meeting,
28 February -2 March. 2007; http://www.gsi.de/documents/DOC-2007-Mar-47-1.pdf. 
\bibitem{CBM_PSD} {\it Guber~F. et al.}// CBM-PSD-note-2006-001. 2006.
\bibitem{MPD_experiment} {\it The multi-purpose detector (MPD) to study heavy ions collisions at NICA}// Letter of Intent. 2008.
%\bibitem{star_K} {\it Adler~C. et al.}// Phys.Lett. 2004. {V.B595.} P.143.
\end{thebibliography}
\end{document}